\documentclass{ws-acs}
\usepackage{color}
\newcommand{\colb}{\color{blue}}
\usepackage[T1]{fontenc}
\usepackage{url}
\begin{document}

\makeatletter
\def\@biblabel#1{[#1]}
\makeatother

\markboth{Andrea Apolloni, Floriana Gargiulo }{Diffusion processes through social groups' dynamics}
\title{ Diffusion processes through social groups' dynamics }
\author{ANDREA APOLLONI}

 \address{Institut des Syst\`emes Complexes Rh\^{o}ne-Alpes (IXXI) and Laboratoire de Physique, \\  \'Ecole Normale Sup\'erieure de Lyon, \\ 69007 Lyon, FRANCE \\
 andrea.apolloni@ens-lyon.fr }

\author{FLORIANA GARGIULO}

       \address{INED\\
       75020, Paris, France \\
       floriana.gargiulo@ined.fr
}

\maketitle
\begin{abstract}
Axelrod's model describes the dissemination of a set of cultural traits in a society constituted by individual agents. In a social context, nevertheless, individual choices toward a specific attitude are also at the basis of the formation of communities, groups and parties. The membership in a group changes completely the behavior of single agents who start acting according to a social identity. Groups act and interact among them as single entities, but still conserve an internal dynamics. 
We show that, under certain conditions of social dynamics, the introduction of group dynamics in a cultural dissemination process avoids the flattening of the culture into a single entity and preserves the multiplicity of cultural attitudes.
We also considered  diffusion processes on this dynamical background, showing the conditions under which information as well as innovation can spread through the population  in a scenario where the groups' choices determine the social structure.
\end{abstract}
\keywords{complex system, groups' dynamics, evolving network, simple diffusion, complex diffusion}
\section{Introduction}
Social groups have been the focus  of many studies in different fields, from  mathematical biology \cite{gueron1995dynamics}, through economics \cite{2009arXiv0904.0761G} and sociology \cite{1987}.
These formations play a fundamental role at many levels, both in understanding the existence of strong ties in  social networks and in throwing light on the diffusion processes mechanism in a population.\\
Two main processes govern the formation of groups in a society: homophily (the tendency to interact with similar agents) and social influence (the tendency to become more similar after an interaction) \cite{1997}. 
Many empirical studies have proven the homophily attraction at different scales: acquaintance networks  \cite{mcpherson01homophily} and voluntary organization \cite{1987} show an homogeneous distribution in respect of some demographic factor; industrial districts are formed by companies sharing the same \emph{local culture} \cite{2009arXiv0904.0761G}. 
The homophily has a double effect on group formation: on one side, it pushes agents to converge to the same opinion; on the other side, it raises "barriers" between individuals of different opinions \cite{centola2007homophily}. 
Strong ties are created among members of the same group, due to their similitude, while weak links are established between members of different groups.
 Agents come together in groups based on their shared opinions and at the same time the agent's membership defines his own opinion \cite{1974}. From this point of view the group's opinion is the individual opinion and viceversa. 
In the context  of network science, a group/community is a set of  tightly clustered individuals;  in this work we are not interested in analyzing internal structure of the groups but rather to deepen the question of the adaptive interactions between groups' structures.
Groups are not static objects but they possess an internal as well as an external dynamic. The external dynamic is related to the interaction with other groups present in the society, while the internal one is related to the individual membership choices.  In this paper we will not explore the choices of individual memberships that govern group dynamics like in \cite{gargiulo2010opinion}, \cite{holme2006nonequilibrium} , but we will deal with a network of groups considering directly the processes that concern these macro-structures.
Two main processes can be pointed out at this level: coalescence and fragmentation.

\emph{Coalescence:} The same homophilous interaction that brings together individuals to form communities is also at the basis of the interaction among different groups. 
Two groups can momentarily align their opinions and then decide to merge. Some examples are electoral coalitions, consortiums of firms to be awarded of a contract, the scientific collaboration between groups to obtain fundings. 
The likelihood of such cooperation depends on the group's open-mindedness: groups, as well as individuals, are likely to interact with similar groups and then become more similar, based on the shared elements. To achieve the merging of two groups, a compromise process to align the opinions is needed; in this sense, in the merging process, it exists as a sort of pay-off for the collaboration. 

\emph{Fragmentation:} The internal dynamics of the groups depends instead on the individual choice to belong or not to the group. This rejection can occur for various reasons, for example, the communication flow inside the group is not working \cite{zachary1977information},  new information is introduced in the group creating discrepancies among the members \cite{Carley1990}, 
some members start developing a new point of view not shared with the majority. These factors,  taken singularly or in combination, bring individuals to maturate the decision of  separating thus leading to the formation of new groups, representing the new opinions.\\

We should notice that   individuals belong to many groups \cite{Groups} or, in general have many interactions outside the main group, but several of these contacts have no influence on their social identities. Furthermore the connection with individuals in other groups increases the information advantage with respect to other members. In a completely closed  group where individuals own the same amount of information and  share all the cultural traits, diffusion of new information is not possible \cite{Kleinberg}. 
 In order to have diffusion a certain degree of heterogeneity in the group is necessary, at least in respect of having acquired or not the token of information. The similarity between groups' member facilitate the communication, as shown in \cite{rogers517diffusion}, but the contacts outside the group (weak ties) are necessary to acquire new information \cite{granovetter1973strength}. The choice of  accepting a message or a novelty depends on other factors that are not at the base of group membership, thus not influencing it: being informed about a gossip does not interfere with political or religious opinion for example.
 In our model we should distinguish between two types of information: one can alter individual membership \cite{Carley1990}; the second one instead not influencing individual membership. The dynamics of how these two types of information spread are considered differently in this work. In the first case, since we don't know the causes that bring individuals to maturate the decision, we  simulate it as a random process. As far as the second case is concerned  the flow of information is due to the contact between members.

The interplay between the coalition and fragmentation tendencies causes abrupt changes in member contact patterns and therefore on the underlying social network. At the same time these simple processes  already reproduce some important properties of the cultural diversity at population level.
It is worth noticing that group dynamics influence the number of different opinions present in a society. on the one hand the coalescence tendency selects the point of view widely spread in the society and tries to lead to a consensus. On the other hand, the birth of new opinions, due to the fragmentation process, gives fuel to new interactions between groups.
The balance between these two tendencies could help a given society avoid flattening to a single opinion.
    
There is a huge body of literature on fragmentation and coalescence dynamics that goes from the pioneering work of Levi in biology \cite{gueron1995dynamics}, to financial group \cite{springerlink:10.1007/978-3-540-69384-0_8} and warfare studies \cite{zhao2009internal}.  
Moreover these studies have considered the dynamics as a totally random process:
groups can merge together independently of their cultural identities, and the fragmentation process can occur at every time with equal probability for each group. In our case we also use a stochastic procedure. In addition to previous models we bias  the coalescence of groups through the cultural similarity and the fragmentation through the group size. We use a vector of opinions instead of  real numbers, in order to stress that the similarity is based not only on the number of common elements but also on the specific ones. 
u
As pointed out in \cite{PhysRevE.81.056107},\cite{gargiulo2010opinion},\cite{holme2006nonequilibrium},\cite{kozma2008consensus},\cite{centola2007homophily}, group's dynamics influences the outcome of diffusion processes inside the society, bursting them inside a group but reducing the possibility of being extended to all the population.
When discussing social diffusion processes, we can consider many different paradigms, depending on the particular phenomena we want to study.
In this paper we examine, without the aim of being exaustive, two different kinds of processes, to give an example of the effect of group structured networks on different spreading phenomena: simple and complex propagation. 
The former is at the basis of diffusion of information, gossip and epidemics.  In this case a single node can trigger the cascade effect; the transmission process is due to a contact between an informed/infected individual  and an uninformed/susceptible individual and can happen with a certain probability.

Complex propagation is related to diffusion processes like the diffusion of innovation \cite{Kleinberg} or the languages competition \cite{abrams2003modelling}. 
In this case, the choice of changing status can be "costly" for an individual and therefore a certain resistance to the process is induced. Individuals deciding whether to switch or not to the opposite status, first of all, compare their situation with their neighbours and then calculate the possible pay-off  of the switching action \cite{Kleinberg} \cite{DamonCentola09032010}.  The authors of \cite{doi:10.1086/521848} claim that this feature can help to understand the reasons why social movements first  build local support and after spread geographically: these movements are risky, requiring a massive participation to become effective and  initially  gain momentum within communities and neighborhoods. After this phase a network of movements at higher scale is created merging different local experiences \cite{diani2003social}.

In both examples, groups represent the places where the transmission mostly happens: the redundancy of links in a group improves the possibility of transmitting information and is necessary for complex propagation.  Since group's dynamic changes abruptly the agents' membership and the contact pattern, it directly affects the diffusion process. 

\begin{figure*}[htbp]
\centering
\includegraphics[scale=0.5]{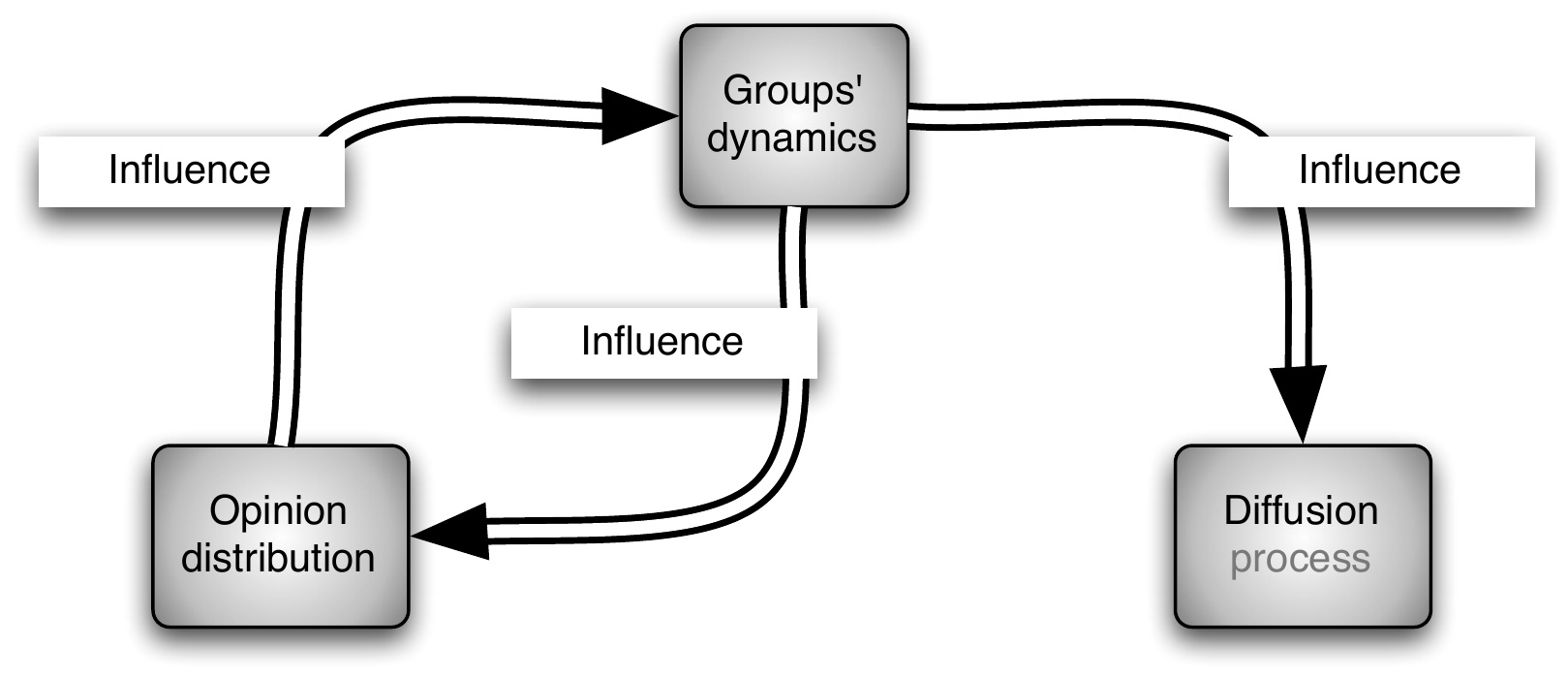}
\caption{The three levels involved in the dynamics. The interaction between groups is mediated through opinions' similarity and this influences the distribution of different opinions in the society. This influences the  opinion dynamics and the groups' dynamics can be seen as co-evolving processes. At the same time, the groups' dynamics could block or enhance the diffusion of gossip or innovation through in a given society. }
\label{3levels}
\end{figure*}

Summarizing, in this paper we discuss a model of fragmentation and coalescence for social groups where these dynamical properties are mediated by the cultural traits of these same structures. The aim is to study the impact of these bulk and intuitive processes on the cultural diversity and on the group decomposition of the society. At the same time  this dynamical scenario will be used  as support for different diffusion phenomena. 
The groups' dynamic is the relevant one, defining at each time the structure and contact pattern between individuals and then the maximum extent of the diffusion. At the same time groups' dynamic influences and is influenced by the opinions present in the population as shown in Figure \ref{3levels}.

In our study we focus our attention on three main aspects of the global scale population:
\begin{itemize}
\item The number of social groups present. 
\item The number of different {\it cultures or opinions} present in the population. Each group is characterized by a social opinion. The coalescence of groups can cause the death, while the fragmentation can cause the birth of new opinions.
\item The final size of simple (for example, gossip) and complex (for example, innovation) diffusion in the population.
\end{itemize}  	
In Section 2 we present the model, defining the parameters ruling the dynamics of the groups and the diffusion processes. In Section 3 we present the numerical results for the group's dynamics and in Section 4 for the diffusion processes. In Section 5  we summarize our conclusions.
In the course of the years many models have been developed to study the political dynamics \cite{springerlink:10.1007/BF01101895} \cite{1999}, where in most of the cases the political system is bi-partited. Our work provides, as in \cite{ISI:000229518900008}, a qualitative model for the case of  a multiparty model dynamics, where  coalitions are formed for election purposes. When  party identities prevail, however, majority could change abruptly  determining  the survival of governments. This is the case, for example, of italian political system  \cite{Giannetti2001529}. The model presented here considers the basic mechanism at the base of the dynamics and we study how majority can be preserved and  if  bi/multi-partisan messages could spread through all the network of political actors.

\section{The model}
\subsection{The endogenous group dynamics}
In this section we describe the endogenous dynamic that concerns the groups' own  identity characterization, namely the cultural aspects that are at the basis of the homophily attraction inside a group.
In our model we define a group as a set of at least 3 agents, sharing the same opinion, namely the social culture.

\begin{figure*}[hbtp]
\centering
\includegraphics[width=8cm]{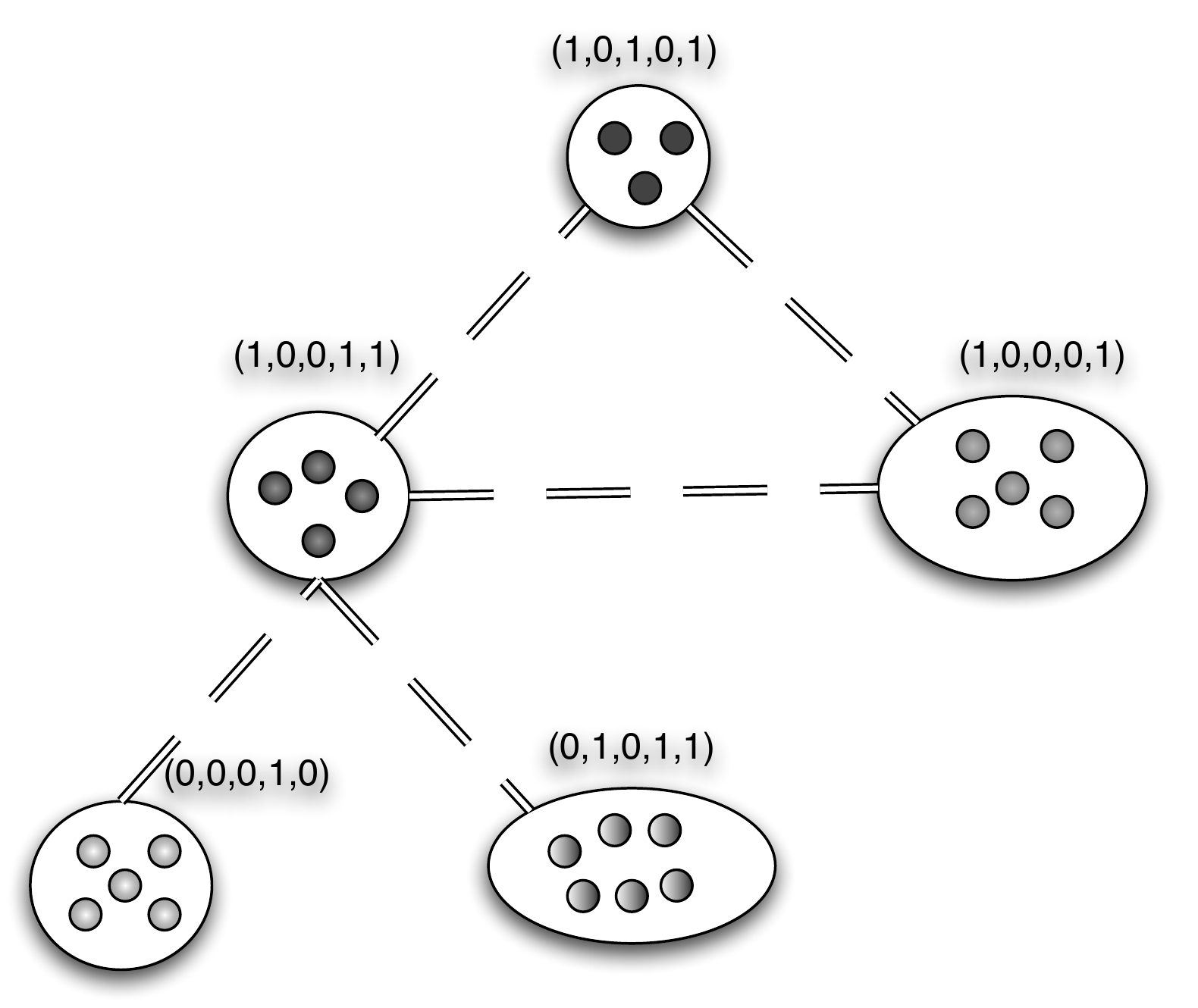}
\caption{Initial groups' distribution. Groups are considered as set of nodes with a bit string representing  groups' opinion.
Groups are connected among them, and can randomly choose with whom interact}
\label{initialdescr}
\end{figure*}

We identify the socio-cultural characterization of a group as a string of binary bits of length $L$; in such a way, we can identify at most $2^L$ different cultural identities. 
Every group $i$ is characterized by its social string, $\phi_{i}$, and by the number of its members, $n_i$.
We can also have simultaneously groups with the same socio-cultural traits, that for their historical pattern or for other reasons (i.e. geographical distance and isolation, other cultural traits not shared) cannot converge into a single entity.
A group can  choose another one with whom  interact, Figure \ref{initialdescr}. If the two groups are culturally similar, they can merge together and  form a larger group.

The cultural similarity between two groups is measured by the similarity function \cite{centola2007homophily}:
\begin{equation}
\Theta(i,j)=\sum_{h=1}^L |\phi_{ih}-\phi_{jh}|
\label{eq1}
 \end{equation}
This function is a `` Hamming distance '' defined as the number of positions for which the corresponding bits differ.
  It ranges between zero and $L$: it assumes the value zero when the two cultural strings  are identical  and $L$ when they have nothing in common.  
The dynamics of the group can allow two different processes: coalescence and fragmentation.  
Coalescence process is mediated by a parameter $\varepsilon$ (called the {\it open-mindedness} parameter), whose value is a real number between 0 and 1 and that represents a threshold for the similarity function. Two groups can merge only if:
\begin{equation}
\Theta(i,j) < L \times \varepsilon
\end{equation}
this means that groups can merge only if they differ for a finite number of elements in the vector. Increasing  $\varepsilon$ means that the differences  between groups are less and less significative for  the fusion, while  diminishing means that groups are more selective in choosing whom to interact with.
If two groups join, the majority group imposes its cultural traits on the smaller one. 
In this sense, we can define an adaptive network of groups: at each time step each group is connected with all the other group structures with whom it can potentially merge, namely the groups whose traits differ less than the open-mindedness parameter.

Simultaneously each group presents a certain tendency to fragment.  This tendency increases with the size of the group:
\begin{equation}
p_{frag}=\frac{\text{(size of the group)}}{N}		
\label{eq2}
 \end{equation}
where $N$ is the total number of agents.  

This dynamical choice is motivated by the following facts. Increasing the size  of the group, the communications inside a group become more difficult  (e.g. Zachary Karate club \cite{zachary1977information})  and this increases the probability that new idea can arise as mutation of pre-existing ones. Furthermore, in the case of firms, since maintaining  relations is costly, a large number of connections can be discarded if no more economically convenient \cite{2009arXiv0904.0761G}. 

If the group splits, a new group is generated {with a new opinion vector, obtained switching one of the trait of the begetting group's one. 
In effect the number of switches/mutations from the begetting group could be considered as an additional parameter. But we observed in the simulations that it is not influencing the model's outcomes.
For sake of consistency with the initial conditions we impose that the minimum size of the group should 3 members.}
 \\
In the simulation process we randomly alternate phases of coalescence and fragmentation of groups: at each step a group randomly decides which process it undergoes. If it decides to join to another group, it chooses randomly the second group and if they are compatible according to the threshold they merge into a new group. If it decides to split, it generates a new group with probability $p_{frag}$. 
The groups' dynamics influences the opinion distribution in the population. While the fragmentation process gives birth to new opinions, the coalescence could determine the  disappearance of certain traits.  According to the particular value of the open-mindedness  parameter $\varepsilon$ some of the traits can survive in the coalescence process, and then  the opinion can partially be transferred to the new group. 
\subsection{Exogeneous dynamical processes}
In the previous section we discussed the group dynamics on the basis of the endogenous processes connected to groups' identity traits. In this paragraph we consider a diffusion process  that has influence on extra dimension (exogenous),  without perturbing group structure, like for example the gossip information or an epidemic spreading. Many examples of  complex diffusion processes can be chosen, the complexity being related on the particular phenomena a modeler wants to describe. The basic idea is to analyze the effect of the group endogenous dynamics on external spreading phenomena.\\
We consider two simple kind of diffusion processes: rumor spreading and  innovation diffusion.
The first type is described as an epidemic model \cite{Nekovee2007457} \cite{Kleinberg} \cite{agliari:046138} \cite{10.1109/CSE.2009.240}:  informed individuals  have acquired a  token of information (Infected) that they can transmit only  to individuals in their social network (Susceptible) with a certain probability. The extent of the diffusion depends on the topology of the underlying network and on the transmission probability. In our case we suppose that an informed individual never forgets the piece of information it has received, or using epidemiological jargon, never recovers as in \cite{agliari:046138} and  \cite{10.1109/CSE.2009.240}.

In this case, the infected individual is an informed one (indicated as individual of type $A$), and the susceptible a not informed one (individual of type $B$). The probability rate of getting informed, given an existing link between the two, is $\beta_A$. The process can be described as a reaction:
\begin{equation}
B+A \xrightarrow{\beta_A} 2A
\label{reaction}
\end{equation}

The diffusion of innovation, as well as the language dynamics, are examples of threshold phenomena: the individual choice to adopt a particular novelty requires simultaneous exposure to multiple  acquaintances that have already adopted it.
From this point of view, this kind of processes  can be seen as a tug of war between the innovators (and adopters), on one side, and the resistant to the novelty, on the other. In this view, the innovation diffusion can be described as a dynamical competition between two species of ideas, the innovative one ($A$) and the conservative one ($B$).  In such competition both transitions are allowed.
An approach that takes into account both the possibility of the innovator and of the conservative to convince each other is the Abrams-Strogatz model for language competition \cite{abrams2003modelling}, \cite{stauffer2007microscopic}. The studied case for Abrams-Strogatz model regards the progressive affirmation of a unique language in a mixed population initially speaking two different idioms. The adoption of a language depends both on the number of individuals that already adopt the language and on the prestige of the idiom. 
We use such approach for describing the diffusion of an innovative idea $A$ in conservative population with idea $B$.  
People can decide to adopt an idea ($A$ or $B$) for direct imitation and according to the number of persons that have already adopted the idea. 
A transmission rate $\beta_A$ is associated to the transition from the conservative to the innovative idea and in this case the process can be describe as in (\ref{reaction}).
At the same time, conservative people offer resistance to the introduction of innovation: conservatives try to convince innovators to go back to the original idea ($B$). The transmission rate in the case 
\begin{equation}
A+B\xrightarrow{\beta_B}2B
\end{equation}
is $\beta_B$.
The  $\beta$ parameters can be thought as a sort of pay-off or perceived prestige.

 In both type of diffusion the seed of the process differs from other member of the groups because of the token of information/novelty he owns. 
The extent of the diffusion depends only on the undergoing network of contact: the piece  of information as well as the novelty, first diffuses in the group and then, due to the coalescence and fragmentation dynamics, and the consequent repartition of the members, can be transmitted to other groups. In this sense we are consdering a mean field approximation, assuming an homogeneous interaction probability among all the members of the same group (strong ties) and uniform null interaction probability outside the group. This is, in effect, a strong assumption, that we considered in order to highlight the direct effect of the endogenous dynamics on the exogenous one.\\
At the beginning of the diffusion there is just one innovator surrounded by a sea of resistant people; the number of people with whom he is in contact depends on the initial partition in groups. \\
To include the group structure inside the diffusion model and to consider stochastic oscillations we used a binomial extraction process inside each single group. 
We consider that the transmission process takes place between agents that are members of the same group. 
Depending if we are considering the diffusion of a piece of information or innovation,  inside each group $i$ a single or a double  contamination mechanism, respectively,  is considered.
Consider the case of simple propagation: at each time step $A_i$ (the number of informed individuals in group  $i$) can increase due to the adoption of the information by some of $B_i$ (the number of non-informed individuals in group  $i$).  Under the assumption of homogeneous mixing inside the group, the probability rate is given by $\beta_AA_i(t)/n_i(t)$. The number of new informed individuals is then a stochastic variable that follows a binomial distribution with the corresponding adoption probability. 
In the case of complex propagation we talk about idea instead of information. The case of diffusion of innovation can be decomposed in two processes happening at the same time in group $i$: novelty adopters increasing their number $A_i$ due to the adoption of information by a resistant; viceversa the number of adopters  $A_i$ could decrease due to the adoption of idea $B$ by some of its member. For the first process the probability rate is the same as for information case, while for the second process is given by $\beta_BB_i(t)/n_i(t)$. The number of new adopters of each idea is then a stochastic variable that follows a binomial distribution with the corresponding adoption probability. The net change of adopter of a specific idea, at time step $t$, is given by the difference between these variables.

Summarizing the entire dynamics (see Figure \ref{dynamics}):
\begin{figure*}[htbp]
\centering
\includegraphics[scale=0.5]{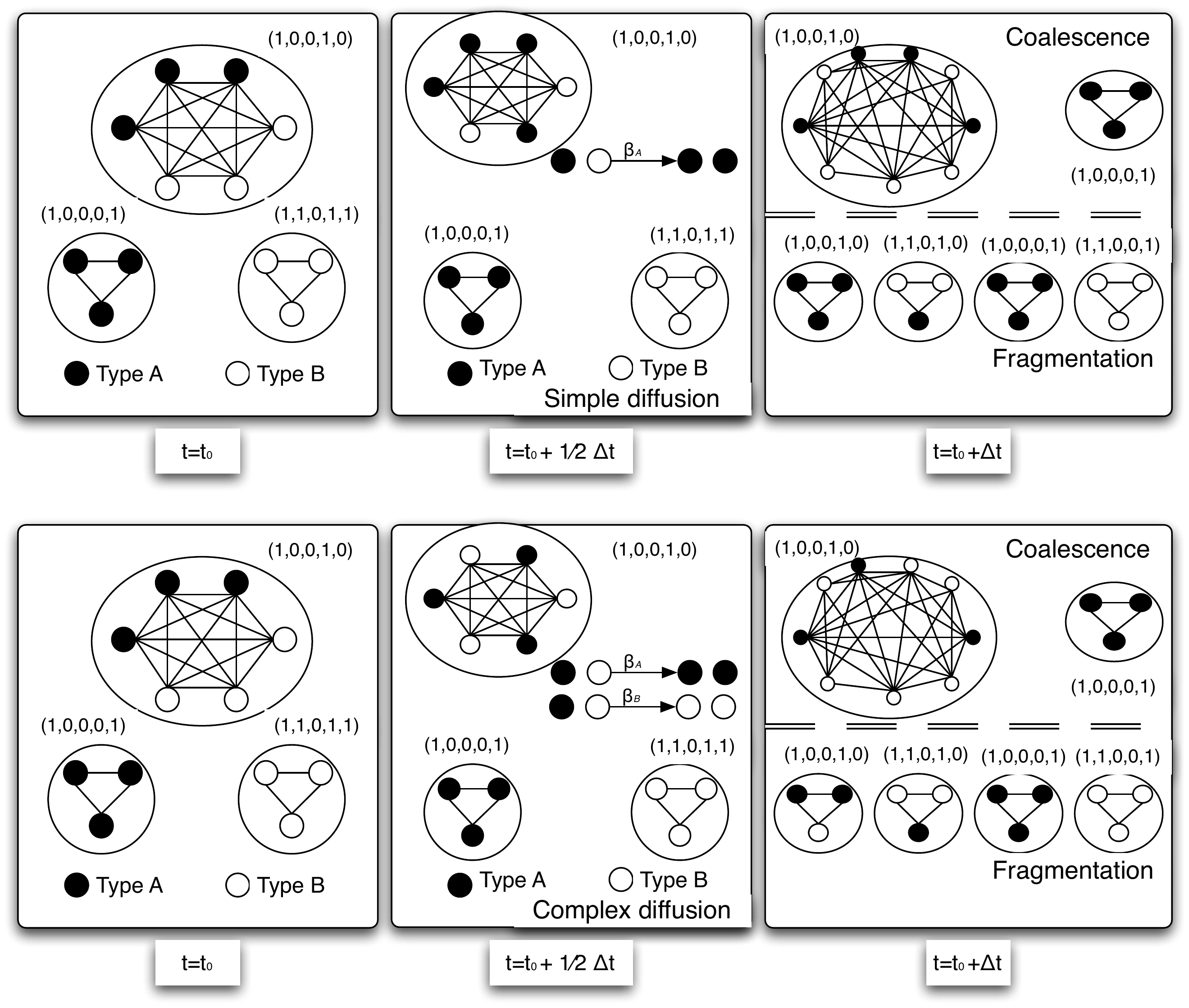}
\caption{Dynamical group contagion model. At each time step  each group is checked. If the number of individuals of type $A$ or $B$ is less than the size of the group, then information/innovation can spread in the group. After the checking, a group is randomly chosen and can either try to join another group, either split in two groups. 
The upper figure  is related to the case of spread of information, the lower one to the case of innovation. In the first case there is a single contagion process, the information flows from  individuals of type $A$ to individuals of type $B$. In the second case a double infectious process is considered: both individuals of type $A$ and $B$ can `` infect" individuals of the other type. }
\label{dynamics}
\end{figure*}

\begin{romanlist}[ii]
\item {\bf Initial condition:} we consider a set of $N$ agents randomly divided in $N_C$ groups, each group endowed with a binary string of length $L$, representing the {\it group's opinion}. All the agents are in state $B$, except one in a randomly chosen group, that is in state $A$.
\item {\bf Diffusion processes:} According to the particular process.
\begin{romanlist}[b]
\item {\it Simple propagation.} Using binomial extraction, the infected agent can infect other agents in the same group, with probability rate $\beta_A$.
\item {\it ComplexPropagation.} We use two binomial extractions. The first one is for the agent in state $A$ infecting  other agents in state $B$ in the group, with probability  rate $\beta_A$. The other one is for agent in state $B$ re-infecting agent in state $A$ with probability rate $\beta_B$.
\end{romanlist}
\item {\bf Groups' dynamics:} We randomly choose a group $i$ and we  perform one of the following actions:
\begin{romanlist}[b]
\item  {\it Coalescence.} Another group $j$ is chosen. The opinions are  compared and if the function $\Theta(i,j)<L \times \varepsilon$ the two groups merge together. The opinion of the resulting group is given by majority rule.
\item {\it Fragmentation.} With probability proportional to the size of the group, the group splits in 2 smaller subgroups. The smaller group's opinion is obtained switching randomly one entry of the vector. Infected agents are randomly distributed in the two groups.
\end{romanlist}
\end{romanlist}
\section{Simulation approach and results - The dynamics of the groups}
In this section we report the results concerning the dynamics of the groups.
We consider a population of size $N=2000$, initially divided in $N_C=100$ groups. 
The agents are randomly assigned to each group at the beginning of the simulation.
We indicate with $n_i$ the size of the $i$-th group.
For all the possible values of $\varepsilon$ we evolved the system for a time T=500, expressed as iterations. To deal with the intrinsic stochasticity of the system, the experiment has been repeated 500 times. The various measures reported in the graphs are obtained as an average of the 500 realizations. 

\begin{figure*}[htbp]
\centering
\includegraphics[scale=0.5]{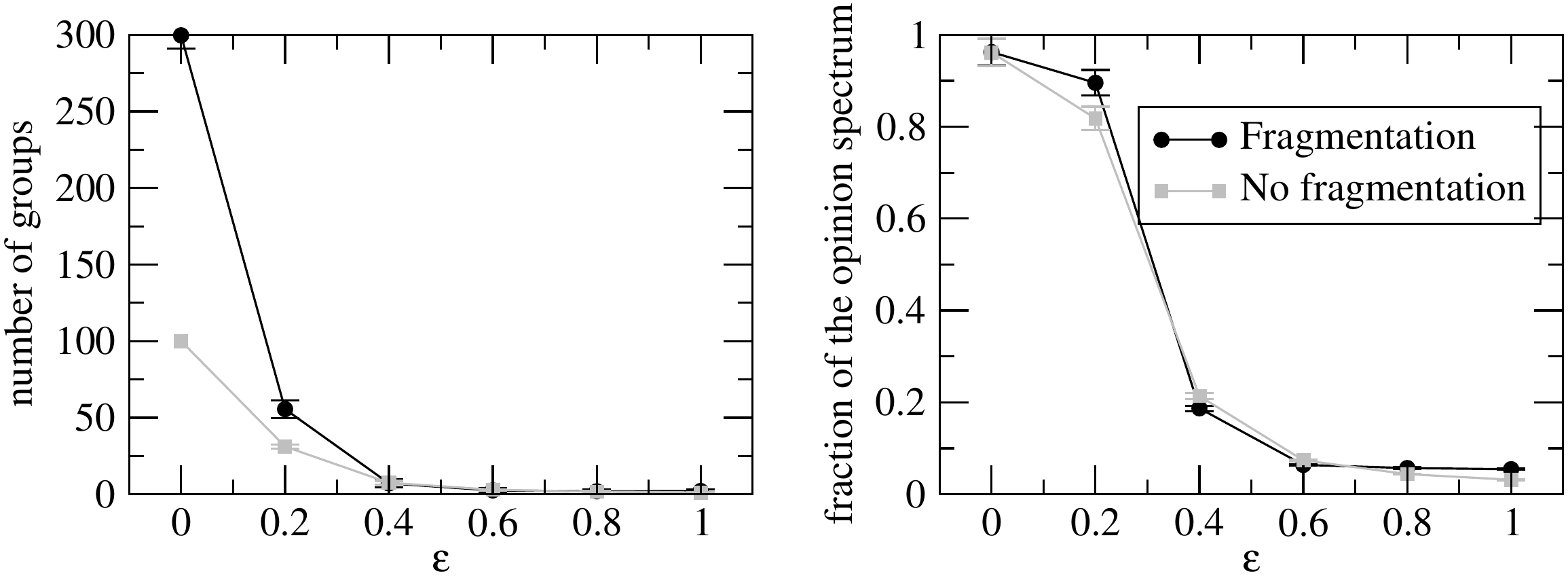}
\caption{Left plot: Final number of groups as a function of the group open-mindedness.  Right plot: Final number of opinion states as a function of the group open-mindedness. 
The circles represent the case where the whole process (merging+fragmentation)  is performed and the squares the case where only merging is considered. The results are averaged over 500 realizations of the system.}
\label{frag-nofrag}
\end{figure*}

The number of groups at the end of the simulation, as well as the opinion diversity, depend on the \emph{open mindedness} parameter $\varepsilon$. Figure  \ref{frag-nofrag} compares two cases: the case where the whole dynamic is considered and the case where the fragmentation process is not performed (groups can only merge).  
 We consider the opinion vector has size $L$=5;
If the open-mindedness parameter is zero, merging is not allowed. Therefore in the case where fragmentation is neither allowed the  number of groups does not change in the simulation.  When the fragmentation is performed, for $\varepsilon=0$, the groups will slowly reach to the maximum possible splitting (considering that a group contains at least 3 members).
As $\varepsilon$ increases, namely the groups become more \emph{tolerant} to the differences, the number of final groups decreases. For $\varepsilon>0.4$ in the case where fragmentation is not performed we observe the formation of a single giant group. In the case where fragmentation is allowed, the situation is not strongly dissimilar. An unstable equilibrium around a giant group is created: at some iteration a new group is generated as a mutation of the giant one and in a second time it is re-absorbed in the giant coalition.  In these case the capacity to generate diversity can not contrast  the inclusive capacity of the aggregation process.
Of course the fact that the number of groups decreases with $\varepsilon$ reflects on the opinion diversity of the population. 
When the giant group is created, only a unique opinion vector, and some mutation of this, can exist. Therefore we observe, for $\varepsilon>0.4$, an almost total consensus formation. 

\begin{figure*}[htbp]
\centering
\includegraphics[scale=0.5]{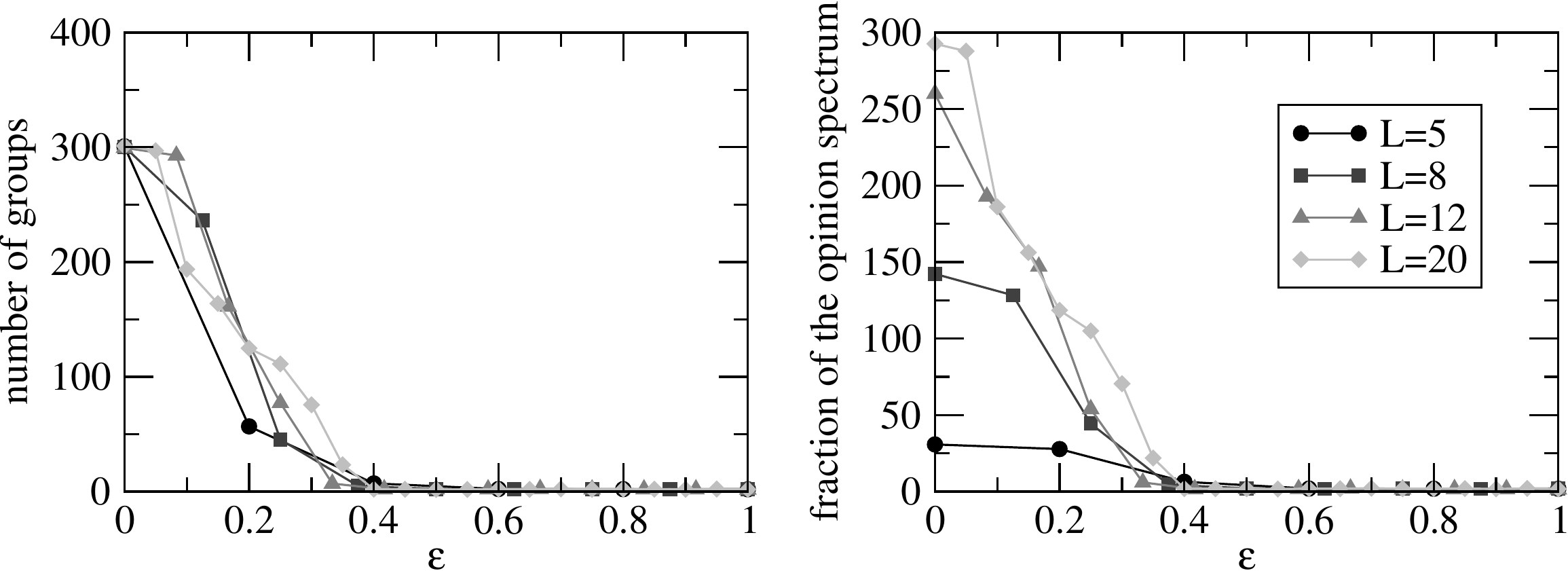}
\caption{Left plot: Final number of groups as a function of the group open-mindedness. Right plot: Final number of opinion states as a function of the group open-mindedness.  
Different type of points correspond to different sizes of the opinion vector (L=5,8,12,20).The results are averaged over 500 realizations of the system.}
\label{L}
\end{figure*}

 Figure \ref{L} shows the final number of groups and opinion in the population as a function of the open-mindedness parameter $\varepsilon$ when varying  the length of the opinion vector $L$ (5,8,12,20). Independently of  the number of possible cultural traits, the system, for $\varepsilon >0.4$, eventually converges to a unique  group. Increasing  the length of the opinion vector , means essentially increasing the number of possible opinion  present in the society. On the other hand, the reduced choice of possible value for each feature increases the possibility of two groups to join, thus, also if  groups share a fraction of feature less than 0.5, consensus in the society can be reached.
 The time to reach the equilibrium changes with $L$ (the convergence is slower when $L$ is larger)  but not enough to motivate deeper analysis.

\section{Simulation approach and results - Diffusion processes on the network}
\subsection{Diffusion of information}
We consider the diffusion of information as gossips while the network is evolving in time  due to the fragmentation and coalescence processes.
In this case we consider that the agents who have been informed can infect only agents belonging to the same group.
The process is simulated using binomial extraction at each time with probability rate $\beta_A$. We have considered five different values for $\beta_A$ ranging form 0.2 to 1.

\begin{figure*}[htbp]
\centering
\includegraphics[scale=0.5]{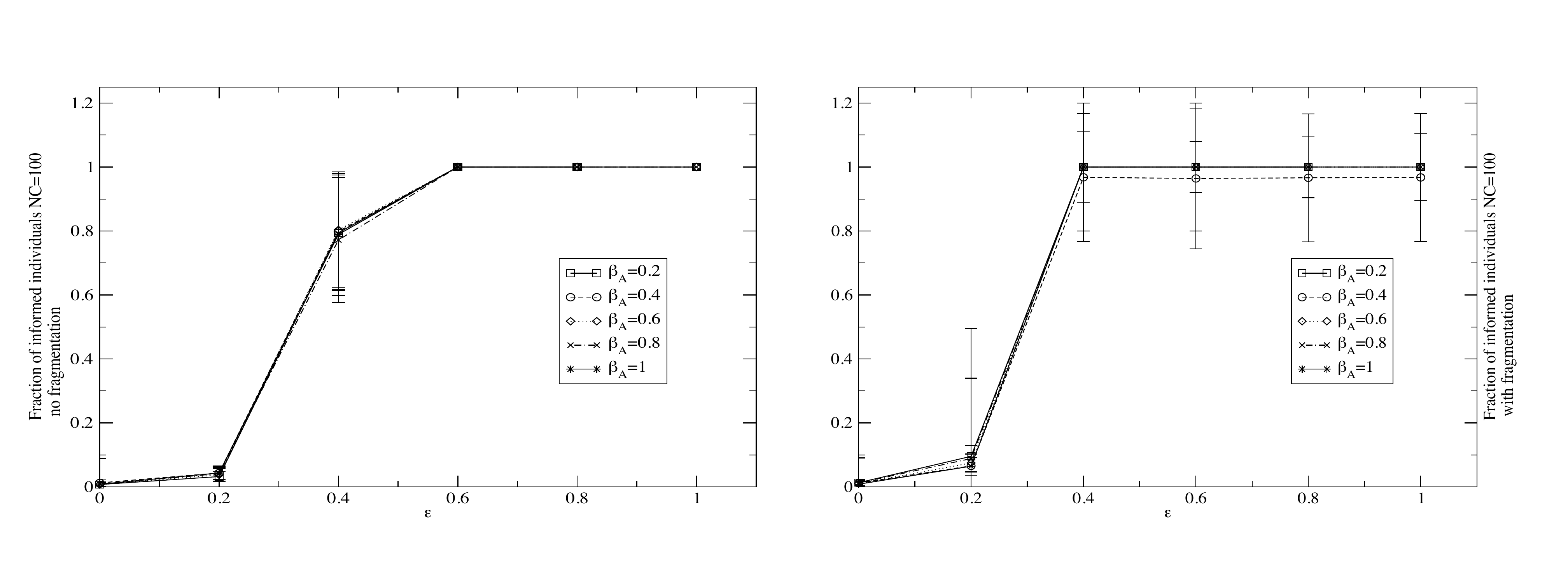}
\caption{ Fraction of individual  informed of a gossip varying $\beta_A$ and $\varepsilon$ with initial number of groups $N_C=100$ and $L=5$. On the left, fragmentation is not performed,  on the right, fragmentation is performed. }
\label{finalgossip}
\end{figure*}

Figure \ref{finalgossip} shows the fraction of individual informed, varying $\beta_A$ and $\varepsilon$  when the population is initially divided in $N_C=100$ groups. On the left we have considered the case when groups can not split but only merge, on the right when merging is allowed.
In the case when groups can not split (figure \ref{finalgossip} on the left)  we notice that the final size is independent of the particular value transmission rate $\beta_A$ but depends on the coalescence process.
 For values of $\varepsilon \geq 0.6$ the gossip spreads through all the network,  for $\varepsilon =0.4$ the group hasn't still merged in an unique one and gossip spreads through a finite part of the network. 

 We notice in this case the the information/rumors has reached a finite fraction of the population, although the extent could vary depending in which group the informant is, and the underlying groups' dynamic.
For smaller values of $\varepsilon$ due to the smallness of the groups' sizes
 the diffusion is restricted to the group where the initially informed is ($\varepsilon=0$) or identical groups ($\varepsilon=0.2$).

On the other hand, the case when $\alpha=1$ and $N_C=100$, we notice that when groups can not merge ($\varepsilon=0$), the information can spread only in the group where the informant is. Due to the smallness of the groups, and  to the fact that the fragmentation process is reducing the sizes, the diffusion can not take off.
When groups with identical opinions can merge together, the fraction of informed individuals  increases. In the end for higher values of the open-mindedness parameter, the gossip can spread through almost all the network independently of the probability rate $\beta_A$. Compare to the previous case, due to the fragmentation process the set of possible scenario is wider. Nevertheless a finite fraction of the population always gets informed.
  
  \subsection{Diffusion of innovation} 
  
\begin{figure*}[htbp]
\centering
  \includegraphics[scale=0.5]{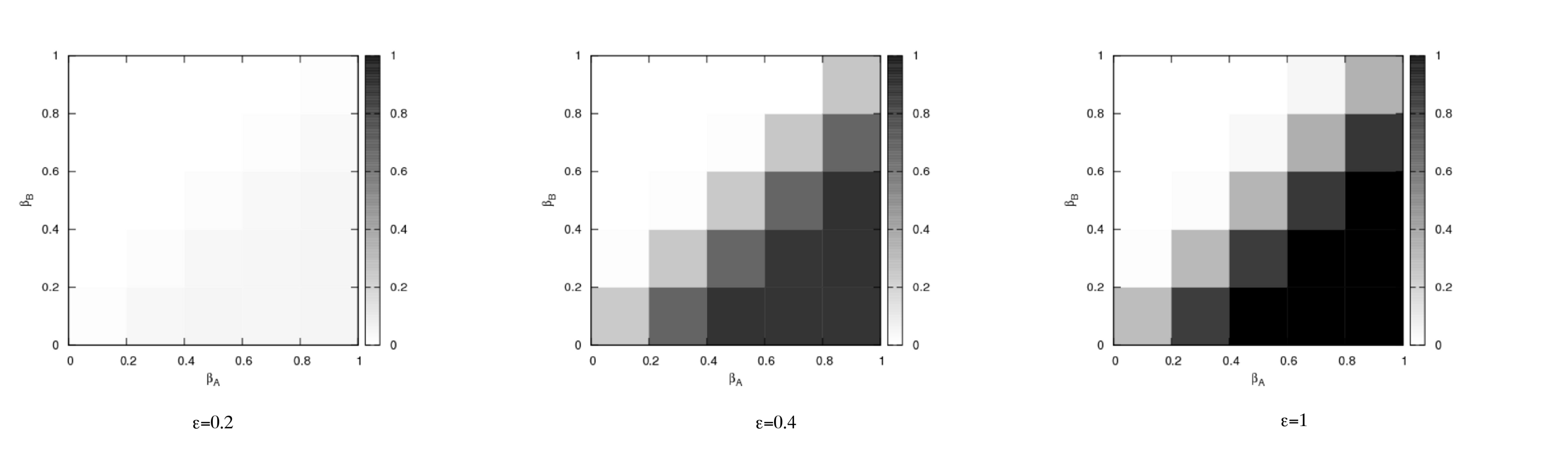}
  \caption{ Fraction of population that has accepted the novelty (type $A$ individuals) when fragmentation is not allowed , varying $\beta_A$ and $\beta_B$. Each plot being evaluated for a specific value of $\varepsilon$}
  \label{finalinfoNC100st}
  \end{figure*}

\begin{figure*}[htbp]
\centering
  \includegraphics[scale=0.5]{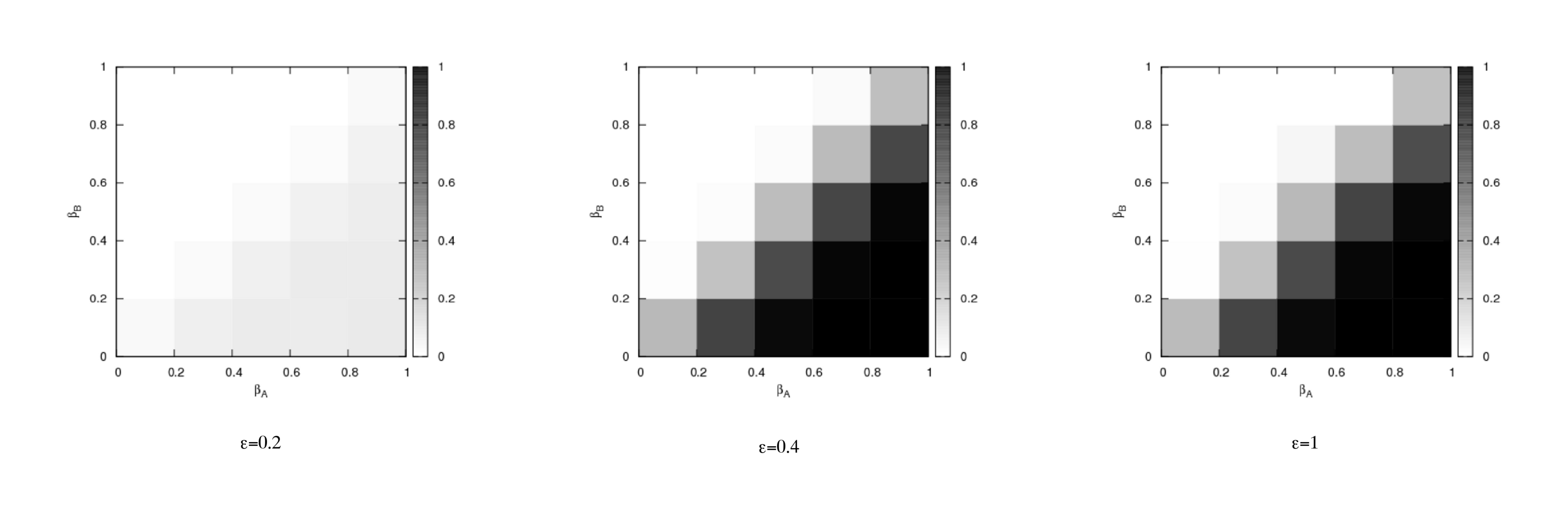}
  \caption{ Fraction of population that has accepted the novelty (type $A$ individuals) when fragmentation is allowed , varying $\beta_A$ and $\beta_B$. Each plot being evaluated for a specific value of $\varepsilon$}  \label{finalinfoNC100dyn}
  \end{figure*}

  We simulate the diffusion of innovation  as a double contamination process with probability rates $\beta_A$ and $\beta_B$, in a population divided initially in $N_C=100$ groups with opinion vector size $L=5$. We consider as extent of the diffusion, the number of agents that at the end of the simulation have accepted innovation (from now on type $A$).
  From this point of view, the diffusion of a gossip can be seen as a particular type as $\beta_B=0$.
  Figures (\ref{finalinfoNC100st}, \ref{finalinfoNC100dyn}) show the  heat maps for final extent at  different values of the parameters $\beta_A,\beta_B$ and $\varepsilon$
 when fragmentation is not allowed (\ref{finalinfoNC100st}), and when is allowed (\ref{finalinfoNC100dyn}). 
  We have reported just the cases significantly different. In fact, for $\varepsilon>0.4$ the innovation has spread through all the population, and the behavior is the same as $\varepsilon=1$. Each plot is evaluated for a specific value of $\varepsilon$ as reported below each figure.  When $\beta_B>\beta_A$  diffusion is not occuring, independently of $\varepsilon$. By contrast, when  $\beta_B<\beta_A$ the range of the diffusion depends  on $\varepsilon$.
  The cases $\varepsilon=0$ is trivial since the diffusion can not take off in both cases: innovator is confined in his own group and can infect only  other members. 
  Increasing $\varepsilon$, the novelty can spread to a finite fraction or to all the population. We notice that for $\varepsilon=0.4$, when  fragmentation is not allowed (\ref{finalinfoNC100st}), a finite fraction of population can be infected, the extent depending on the stochasticity of the diffusion and the  groups' dynamics processes. In the corresponding case when fragmentation is allowed (\ref{finalinfoNC100dyn}), the average fraction of population that has acquired the novelty is larger, almost the totality.  
 For $\varepsilon>0.4$ the novelty spreads through all the population. 
 In all the cases, when $\varepsilon=0.4$ the possible scenarios are wider,since the extension depends on the underlying groups' dynamic.  

\section{Conclusions}
We presented a model of groups' dynamic where the interaction among groups is mediated by the open-mindedness of the groups. 
At the same time we have studied diffusion of gossip and innovation through the population, representing examples of simple and complex propagation.
We have particularly focused on the effect of the open-mindedness parameter on all the processes.
We have found that independently on 
the length of the vector opinion, when the open-mindedness parameter $\varepsilon$ is larger than the value 0.4, groups merge together forming a unique group. 
Two points are worthy of note: firstly, the Hamming distance that we have used for defining the similarity depends not only on the number of common entries, but also on their position. This means that there could be merging only if specific cultural traits are equal. 
Secondly, $\epsilon$ represents a fraction of different elements. Increasing  the size of the vector $L$, we add more cultural traits, but when  less than 40$\%$ of them are different, it is possible to reach a large (in most cases unique) consensus in the society. 

When studying the simple diffusion process, for example gossip, the  fragmentation process plays a double role: when the open-mindedness parameter is low than $\varepsilon=0.4$ it restricts the diffusion to the groups where the initial informed is; on the other hand, when groups are allowed to merge, the fragmentation bursts the process. This is mainly due to the fact that in the population under examination almost all the possible opinions are present and new groups, created through fragmentation, can easily merge with already present.
This cannot happen in the case when fragmentation is not performed. 

Moreover the effect of fragmentation can be seen also in the case of the diffusion of innovation. In the regime where the conservative probability rate $\beta_B$ is larger than the innovator one $\beta_A$ the innovation cannot spread, independently of the open-mindedness and if the fragmentation is performed.
Conversely, in the case where the fragmentation is allowed, the extent of the diffusion depends on the open-mindedness: when groups are "tolerant" with respect to differences in cultural traits, the innovation can spread through all the population; in the case of selective or non-interacting groups, the diffusion cannot spread, but remained confined in the innovator's group or groups with identical opinion.
Since new groups cannot be created that could act as intermediate, the extent of the innovation diffusion cannot be all the population in the case that fragmentation is not allowed and $\varepsilon=0.4$.

\section{Acknowledgments}
A.A. work is supported by DynaNets. DynaNets acknowledges the financial support of the Future and Emerging Technologies (FET) program within the Seventh Framework Program for Research of the European Commission, under FET-Open grant number: 233847
\bibliographystyle{ws-acs}
\bibliography{ws-acs}
\end{document}